\documentclass[a4paper,12pt]{article}

\usepackage{graphicx}
\usepackage{amsmath}
\usepackage{multirow}
\usepackage{float} 

\begin{document}

\begin{flushright}
preprint SHEP-11-12\\
\today
\end{flushright}
\vspace*{1.0truecm}

\begin{center}
{\large\bf Very Light CP-odd Higgs bosons of the NMSSM at the LHC in $4b$-quark final states}\\
\vspace*{1.0truecm}
{\large M. M. Almarashi and S. Moretti}\\
\vspace*{0.5truecm}
{\it School of Physics \& Astronomy, \\
 University of Southampton, Southampton, SO17 1BJ, UK}
\end{center}

\vspace*{1.0truecm}
\begin{center}
\begin{abstract}
\noindent 
We study the detectability of the lightest CP-odd Higgs boson of the NMSSM, $a_1$, 
at the LHC through its production in association with a bottom-quark pair followed by the 
$a_1\to b\bar b$ decay. It is shown that, for large $\tan\beta$ and very high luminosity
 of the LHC, there exist regions of the NMSSM parameter space that can be
exploited to detect the $a_1$ through this channel. This signature is a characteristic 
feature of the NMSSM in comparison to the MSSM,
as $a_1$ masses involved are well below those allowed in the MSSM
for the corresponding CP-odd Higgs state. 

\end{abstract}
\end{center}

\section{Introduction}
Supersymmetry (SUSY) is one of the preferred candidates for physics beyond the Standard Model (SM).
The simplest version of SUSY is the Minimal Supersymmetric Standard Model (MSSM). This realisation however suffers
from two critical flaws: the $\mu$-problem and the little hierarchy problem. The former one results from
the fact that the superpotential has a dimensional parameter, $\mu$
(the so-called `Higgs(ino) mass parameter'), that, due to SUSY, its natural values would be 
either 0 or the Plank mass scale; yet, phenomenologically,
in order to achieve Electro-Weak Symmetry Breaking (EWSB),
it is required to take values of order of the EW scale, 100 GeV, or, possibly, up to the TeV range.
 The latter one emerges from LEP, which failed to detect a light CP-even Higgs boson, thereby imposing 
severe constraints on a SM-like Higgs boson mass, $114$ GeV being its lower limit from data, thereby 
requiring unnaturally large higher order corrections from both the SM and SUSY particle spectrum 
(chiefly, from third generation quarks and squarks) in order to pass such experimental constraints 
(recall that at tree level the lightest CP-even Higgs boson mass of the MSSM is less than $M_Z$).

The simplest SUSY realisation beyond the MSSM which can solve these two problems at once is the Next-to-Minimal Supersymmetric
Standard Model (NMSSM) \cite{review}. This model includes a singlet superfield in addition to the two MSSM-type Higgs
doublets, giving rise to seven Higgs bosons: three CP-even Higgses $h_{1, 2, 3}$ ($m_{h_1} < m_{h_2} < m_{h_3}$), 
two CP-odd Higgses $a_{1, 2}$ ($m_{a_1} < m_{a_2} $) and a pair of charged Higgses $h^{\pm}$. When the scalar
component of the singlet superfield acquires a Vacuum Expectation Value (VEV), an `effective' $\mu$-term, $\mu_{\rm eff}$, will be
automatically generated and can rather naturally have values of order of the EW/TeV scale, as required. Moreover,
the NMSSM can solve the little hierarchy problem as well, or at least alleviate it greatly,
 in two ways: firstly, a SM-like Higgs boson can unconventionally decay into two $a_{1}$'s with $m_{a_1}<2m_b$ \cite{excess}, thus 
avoiding current Higgs bounds (yet this mass
region is highly constrained by ALEPH \cite{Schael:2010aw}
and BaBar \cite{Aubert:2009cka} data); secondly, a CP-even Higgs ($h_1$ or $h_2$) has naturally reduced couplings to the $Z$ boson
due to the additional Higgs singlet field of the NMSSM and the ensuing mixing with the Higgs doublets.

One of the primary  goals of present and future colliders is looking for Higgs bosons. In regard to the Higgs
sector of the NMSSM, there has been some work dedicated to explore the detectability of at least one Higgs boson at
the Large Hadron Collider (LHC) and the Tevatron. In particular, some efforts have been made to extend 
the so-called `no-lose theorem' of the MSSM --
stating that at least one Higgs boson of the MSSM should be found at the LHC via the usual SM-like production
and decay channels throughout the entire MSSM parameter space \cite{NoLoseMSSM} -- to the case of the NMSSM 
\cite{NMSSM-Points,NoLoseNMSSM1,Shobig2}. By assuming that 
Higgs-to-Higgs decays are not allowed, it was realised that
at least one Higgs boson of the NMSSM will be discovered at the LHC. However, this theorem could be violated
if Higgs-to-SUSY particle decays are kinematically allowed (e.g., into neutralino pairs, yielding invisible Higgs signals) 
\cite{Cyril,NMSSM-Benchmarks}.

 So far, there is no conclusive evidence
that the `no-lose theorem' can be confirmed in the context of the NMSSM. In order to establish the theorem for the NMSSM, 
Higgs-to-Higgs decay should be taken into account, in particular $h_1\to a_1a_1$. Such a decay can in fact be dominant in large 
regions of NMSSM parameter space, for instance, for small $A_k$ \cite{Almarashi:2010jm},  and may not give Higgs 
signals with sufficient significance at the LHC. 

Besides, there have also been some attempts to distinguish the NMSSM Higgs sector from the MSSM one, by affirming a so-called 
`more-to-gain theorem' 
\cite{Shobig1,Erice,CPNSH,Almarashi:2010jm,Almarashi:2011hj}. That is, to assess whether 
there exist some areas of the NMSSM parameter space where more and/or different Higgs bosons can be discovered at 
the LHC compared with what is expected from the MSSM. Some comparisons between NMSSM and
MSSM phenomenology, specifically in the Higgs sectors of the two
SUSY realisations, can be found in \cite{Mahmoudi:2010xp}.

In this analysis, we explore the two theorems at once through studying the direct production of a very
 light $a_1$ (with $m_{a_1}<M_Z$) in association with
$b$-quark pairs followed by the $a_1\to b\bar b$ decay, hence a 
$4b$-quark final state,  at the LHC. This channel has a very large cross section at large $\tan\beta$ yet, being 
a totally hadronic signal, is plagued by very large (both reducible and irreducible) backgrounds.

 This work is complementary to the one carried in 
\cite{Almarashi:2010jm,Almarashi:2011hj}, 
in which we explored the $\tau^+\tau^-$, $\gamma\gamma$ and $\mu^+\mu^-$ decay modes of such a light $a_1$ state (again, produced
in association with $b$-quark pairs).

This paper is organised as follows: in Sec. 2, we describe the parameter space scan performed and give inclusive event rates
for the signal. In Sec. 3, we analyse signal and QCD backgrounds for some benchmark points. Finally, we summarise and conclude
in Sect. 4.\\

\section{\large Parameter Scan and Inclusive Signal Rates}
\label{sect:rates}

In our exploration of the Higgs sector of the NMSSM,
we used here the fortran package NMSSMTools developed in 
Refs.~\cite{NMHDECAY,NMSSMTools}. This package
computes the masses, couplings and decay rates (widths 
and Branching Ratios (BRs)) of all the Higgs
bosons of the NMSSM in terms of its parameters at the EW scale.  NMSSMTools also takes into account 
theoretical as well as experimental constraints from negative Higgs searches at LEP \cite{LEP} and 
the Tevatron\footnote{Speculations
of an excess at LEP which could be attributed to NMSSM Higgs bosons are found in \cite{excess}.},
along with the unconventional channels relevant for the NMSSM.

The features of the scan performed have been already discussed in
Refs.~\cite{Almarashi:2010jm,Almarashi:2011hj}, to which we refer the reader for details.
We map the NMSSM parameter space in terms of six
independent input quantities: the Yukawa couplings $\lambda$ and
$\kappa$, the soft trilinear terms $A_\lambda$ and $A_\kappa$, plus 
tan$\beta$ (the ratio of the VEVs of the two Higgs doublets) and $\mu_{\rm eff} = \lambda\langle S\rangle$
(where $\langle S\rangle$ is the VEV of the Higgs singlet). 

For successful data points generated in the scan, i.e., those that
pass both theoretical and experimental constraints, 
we used CalcHEP \cite{CalcHEP} to determine the
cross-sections for NMSSM Higgs production\footnote{We adopt herein
CTEQ6L \cite{cteq} as parton distribution functions, with scale $Q=\sqrt{\hat{s}}$, the centre-of-mass energy
at parton level, for all processes computed.}. As the SUSY mass scales 
have been arbitrarily set well above the EW one (see Refs.~\cite{Almarashi:2010jm,Almarashi:2011hj}), 
the NMSSM Higgs production modes
exploitable in simulations at the LHC are those involving couplings to
heavy ordinary matter only. Amongst the production channels onset by the latter, we focus here on 
$
gg,q\bar q\to b\bar b~{a_1},
$
i.e., Higgs production in association with a $b$-quark pair. This production mode is the dominant one at 
large $\tan\beta$.
We chose $gg,q\bar q\to b\bar ba_1\to b\bar b b\bar b$ also because $b$-tagging can be exploited to trigger 
on the signal and enable us to require four displaced vertices in order to
reject light jets.  The ensuing $4b$ signature (in which we do not
enforce a charge measurement) has already been exploited to detect neutral Higgs bosons of the
 MSSM at the LHC and proved useful, provided $\tan\beta$ is large and the collider has good 
efficiency and purity in tagging $b$-quark jets, albeit
for the case of rather heavy Higgs states (with masses beyond $M_Z$, typically)
\cite{Dai:1994vu,Dai:1996rn}.

As an initial step of the analysis, we have computed the fully inclusive signal production 
cross-section times the decay BR  against each of
the six parameters of the NMSSM. Figs. 1 and 2 present the results of our scan, the first series of
plots (in Fig. 1) illustrating the distribution
of event rates over the six NMSSM parameters plus as a function of the BR and  of $m_{a_1}$. 
The plots in Fig. 2 display instead the correlations between  the
 $a_1\to b\bar b$ decay rate versus the $a_1$ mass
and the $a_1\to\gamma\gamma$ decay rate. It is clear from Fig. 1 that, for our parameter space, 
the large $\tan\beta$ and small $\mu_{\rm eff}$ (and, to some extent, also small $\lambda$) region is the one most 
compatible with current theoretical and experimental constraints, while the distributions in $\kappa$,
$A_\lambda$ and $A_\kappa$ are rather uniform (top six panes in Fig. 1). From a close look at the bottom-left pane
of Fig. 1, it is further clear 
that the BR$(a_1\to b\bar b)$ is dominant for most points in the parameter space, about 90\% and above.
In addition, by looking at the the bottom-right pane of Fig. 1, it is remarkable to notice that the event rates are sizable
in most regions of parameter space, topping the
$10^7$ fb level for small values of $m_{a_1}$ and are decreasing rapidly with increasing $m_{a_1}$. However, there are
some points in parameter space, with $m_{a_1}$ between 40 to 120 GeV, as shown in the left pane of Fig. 2,
 in which the BR$(a_1\to b\bar b)$ is reduced due to the enhancement of the BR($a_1\to \gamma\gamma$) 
(see right pane of the same figure), phenomenon peculiar to the NMSSM and which depends upon the amount
of Higgs singlet-doublet mixing, see \cite{Almarashi:2010jm}. 

\section{Signal-to-Background Analysis}
\label{sect:S2B}

We perform here a partonic signal-to-background ($S/B$) analysis, based on CalcHEP results. We assume $\sqrt s=14$ TeV 
throughout for the LHC energy. We apply the following cuts in our calculations: 
\begin{equation}
\Delta R (j,j)> 0.4,\qquad
\arrowvert\eta(j)\arrowvert<2.5,\qquad
P_{T}(j)>15~{\rm{GeV}}.
\label{cuts:BB}
\end{equation}
Here, we assume that the $b$-tagging probability of a $b$-quark 
is 50\% while the mis-tagging probability of a gluon is 1\% and the one for 
a light quark is 2\%. Figs. 3-7 show the distributions of
the invariant mass of a two $b$-jet (di-jet) system after multiplying the production times decay rates (after cuts)
by the aforementioned efficiency/rejection factors, i.e., true $b$-tagging and mis-tagging probability.
It is clear that the largest background is the irreducible one $b\bar bb\bar b$, which is
one order of magnitude larger than the reducible background $b\bar bgg$. Further, the other reducible
background, involving light quarks, labelled here as $b\bar bc\bar c$, is negligible compared to the 
other two. Notice that all these backgrounds reach their maximum in invariant mass at around 40 GeV.
 Our plots have a bin width of 1 GeV and account for all
combinatorial effects (as appropriate in absence of jet-charge determination).

To claim discovery of the lightest CP-odd Higgs, $a_1$, at the LHC, we plotted in Fig. 8 both signal significances,
left-panes, and corresponding signal event rates, right-panes, as a function of the collected luminosity, and after integrating over
10 and 5 bins (thus mimicking a more optimistic and a more
conservative, respectively, detector resolution in mass of a di-jet system). From this figure, there is a small
window to detect $a_1$ with masses between 35 at 50 GeV at a very large luminosity, $L=300$ $fb^{-1}$, which could well pertain to the
final number of recorded events of the LHC at design luminosity. Further, for an upgraded LHC, with a tenfold 
increase in design luminosity, 
known as the Super-LHC (SLHC), which could well lead to data sample 
with $L=1000$ $fb^{-1}$, there is clear potential to detect an $a_1$ with mass up to $80$ GeV or so. 
This is the more probable the better
the experimental resolution, naturally. 

Finally, notice in Figs. 9-10 the distributions in 
pseudorapidity-azimuth (the standard cone measure) separation of di-jet pairs and average transverse
momentum of the jets, respectively
(in the latter case we also distinguish, in the reducible backgrounds, between $b$-jets and non-$b$-jets, by exploiting 
our knowledge of the Monte Carlo `truth'). Guided by Refs. 
\cite{Dai:1994vu,Dai:1996rn}, this has been done to
check whether further cuts (in addition to those in 
eq.~(\ref{cuts:BB})) could be employed  to improve the chances of detecting these particular final states. While some
combinations of cuts in these
quantities are more efficient for the signal than the backgrounds, there is no overall gain in the signal
significance, further, to the cost of a reduced signal absolute rate. (Results are here shown
for just one signal mass value, yet they are typical also for the other masses considered too.)

\section{Conclusions}
\label{sect:summa}
The Higgs sector of the NMSSM is phenomenologically richer than the one of the MSSM as it has two
more (neutral) Higgs states. In this paper, we explored the detectability of the lightest 
CP-odd Higgs state of the NMSSM, $a_1$, at the LHC and SLHC, through its production in
association with a $b\bar b$ pair followed by $a_1\to b\bar b$ decay. We have shown
that there are some regions of NMSSM parameter space where a rather light $a_1$ state can be discovered at very large
integrated luminosity using standard reconstruction techniques 
\cite{ATLAS-TDR,CMS-TDR}. This is true so long that $\tan\beta$ is very large, typically above 30, 
over an $m_{a_1}$ interval extending from 20 GeV to $80$ GeV,
 depending on the collider configuration. High $b$-tagging efficiency is crucial 
to achieve this, to transverse momenta as low as 15 GeV, ideally,
also accompanied by di-jet mass resolutions up to 10 GeV.
 
Firstly, these results (together with those of 
Refs.\cite{Almarashi:2010jm,Almarashi:2011hj}) support the `no-lose theorem' by looking for the (quite possibly resolvable) 
direct production of a light $a_1$ rather than
looking for its production through the decay $h_{1, 2}\to a_1a_1$,
whose detectability remains uncertain. Secondly, they enable distinguishing the NMSSM Higgs
sector from MSSM one (thus contributing to enforcing a `more-to-gain theorem') since such light CP-odd Higgs states,
below $M_Z$ (which have a large singlet
component), are not at all possible in the context of the MSSM.\\
In reaching these conclusions we have, first, sampled the entire parameter space of the NMSSM to extract 
the regions which are interesting phenomenologically and, then, sampled five benchmark points in the NMSSM parameter 
space (see the captions to Figs. 3-7), all taken at large $\tan\beta$, for which we have extracted the detector 
performances and collider luminosities required for discovery. The encouraging results obtained should help
to motivate further and more detailed analyses.

\section*{Acknowledgments}
This work is 
supported in part by the NExT Institute. M. M. A. acknowledges
a scholarship granted to him by Taibah University (Saudi Arabia).

\begin{figure}
 \centering\begin{tabular}{cc}
    \includegraphics[scale=0.60]{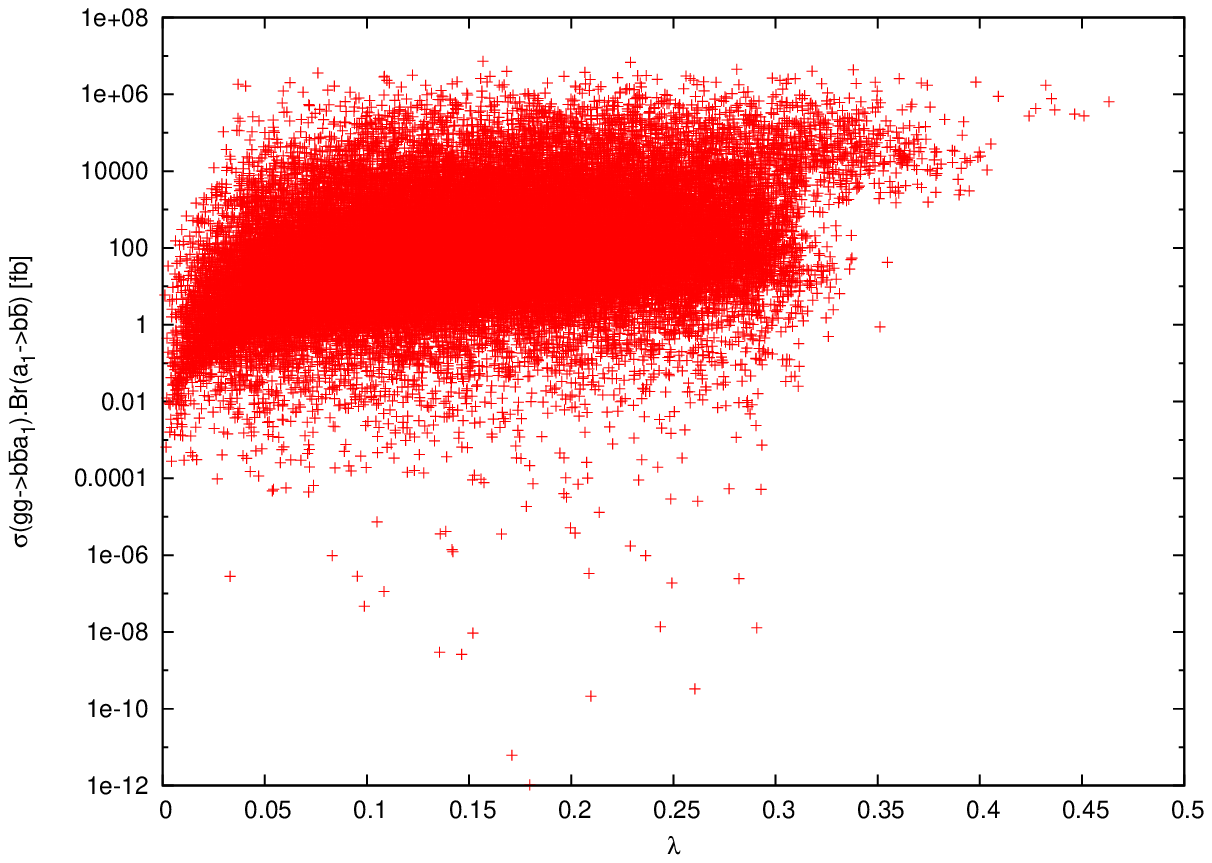}&\includegraphics[scale=0.60]{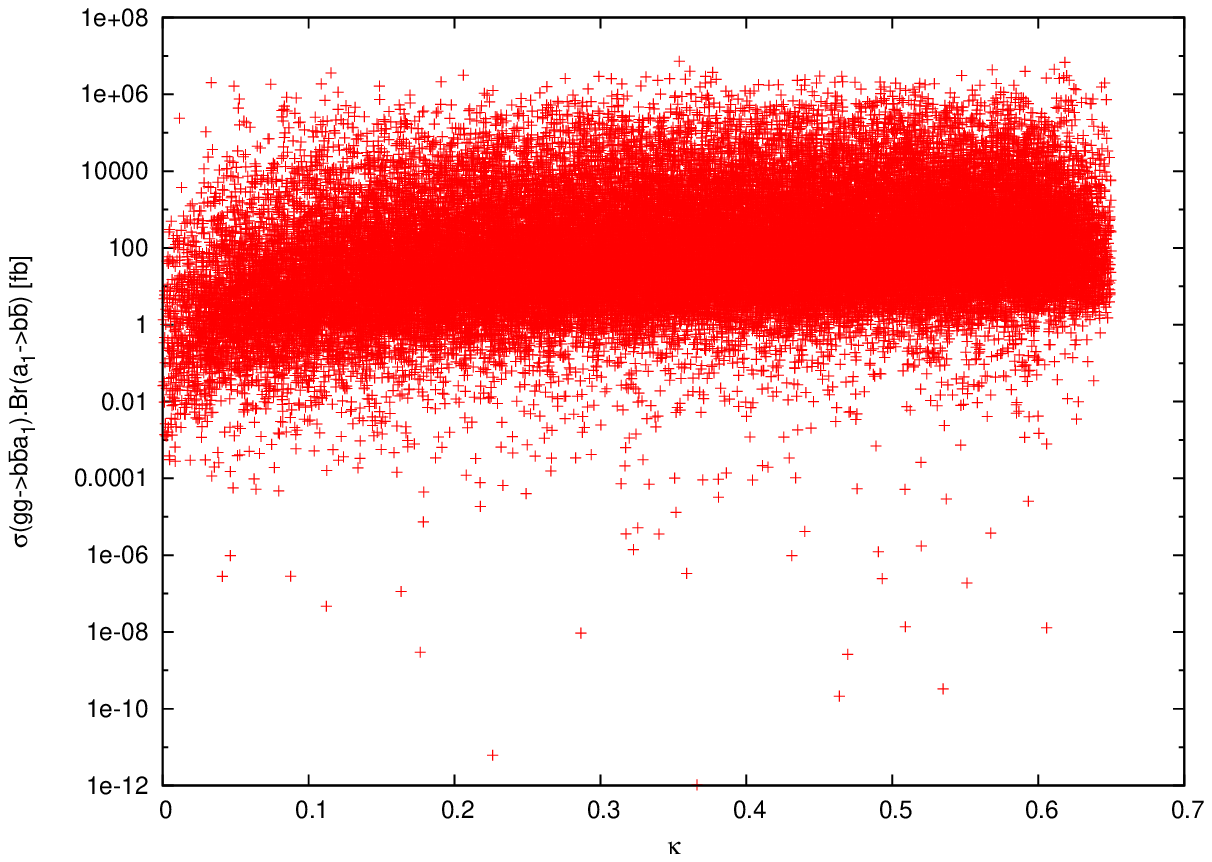}\\
     \includegraphics[scale=0.6]{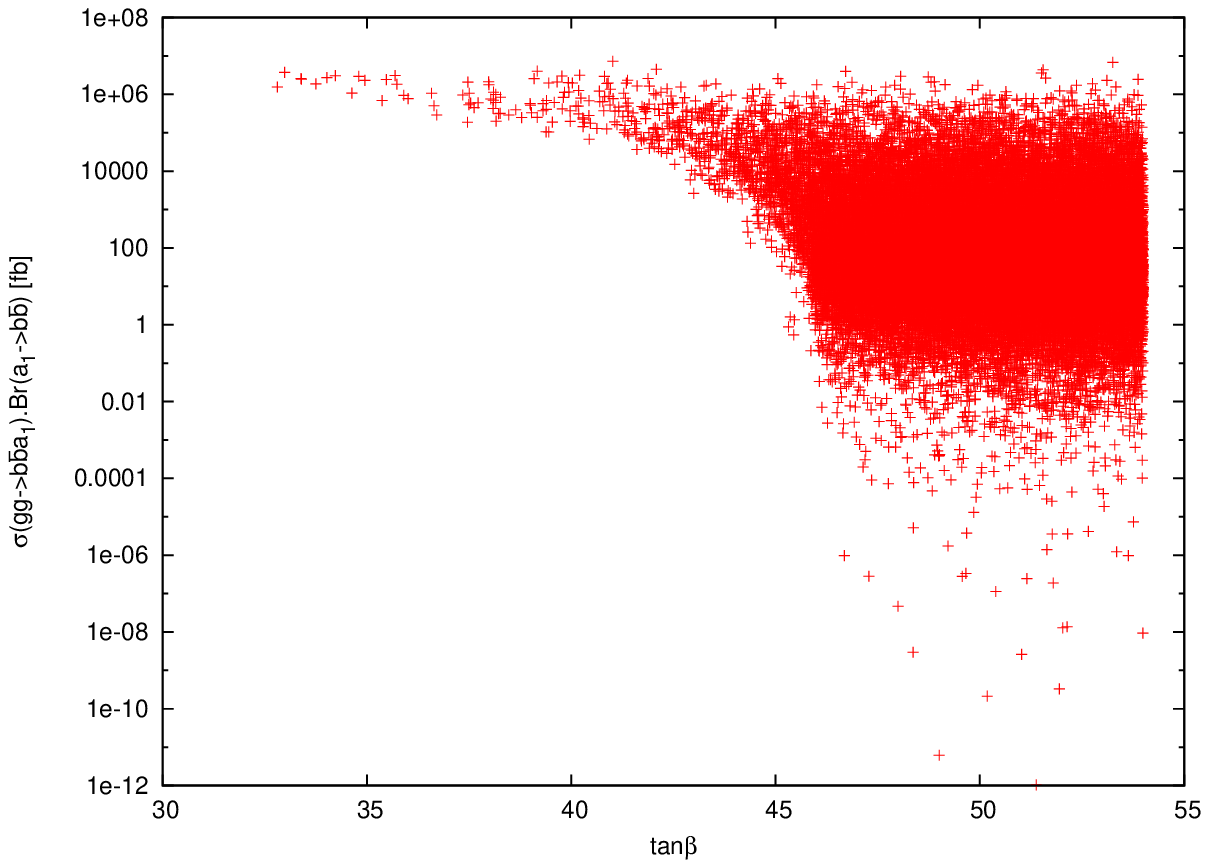}&\includegraphics[scale=0.60]{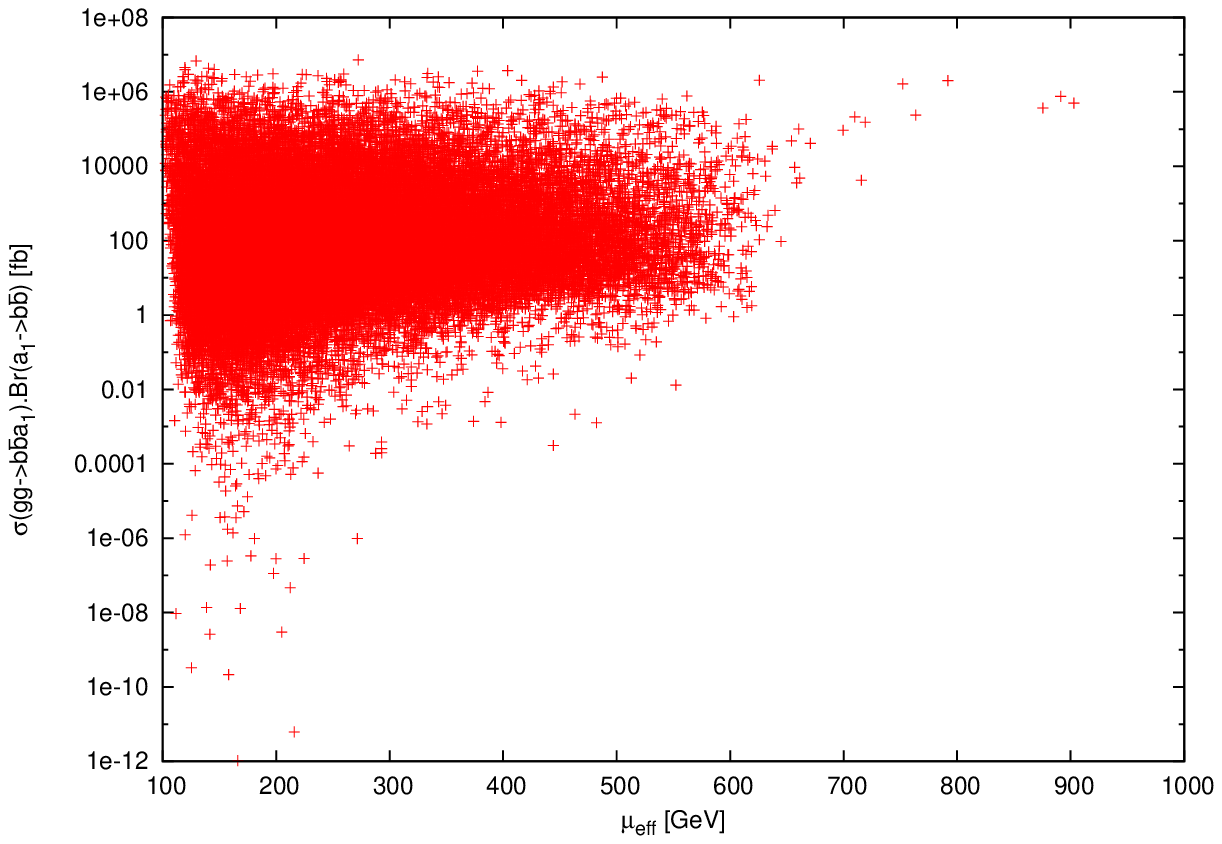}\\
     \includegraphics[scale=0.60]{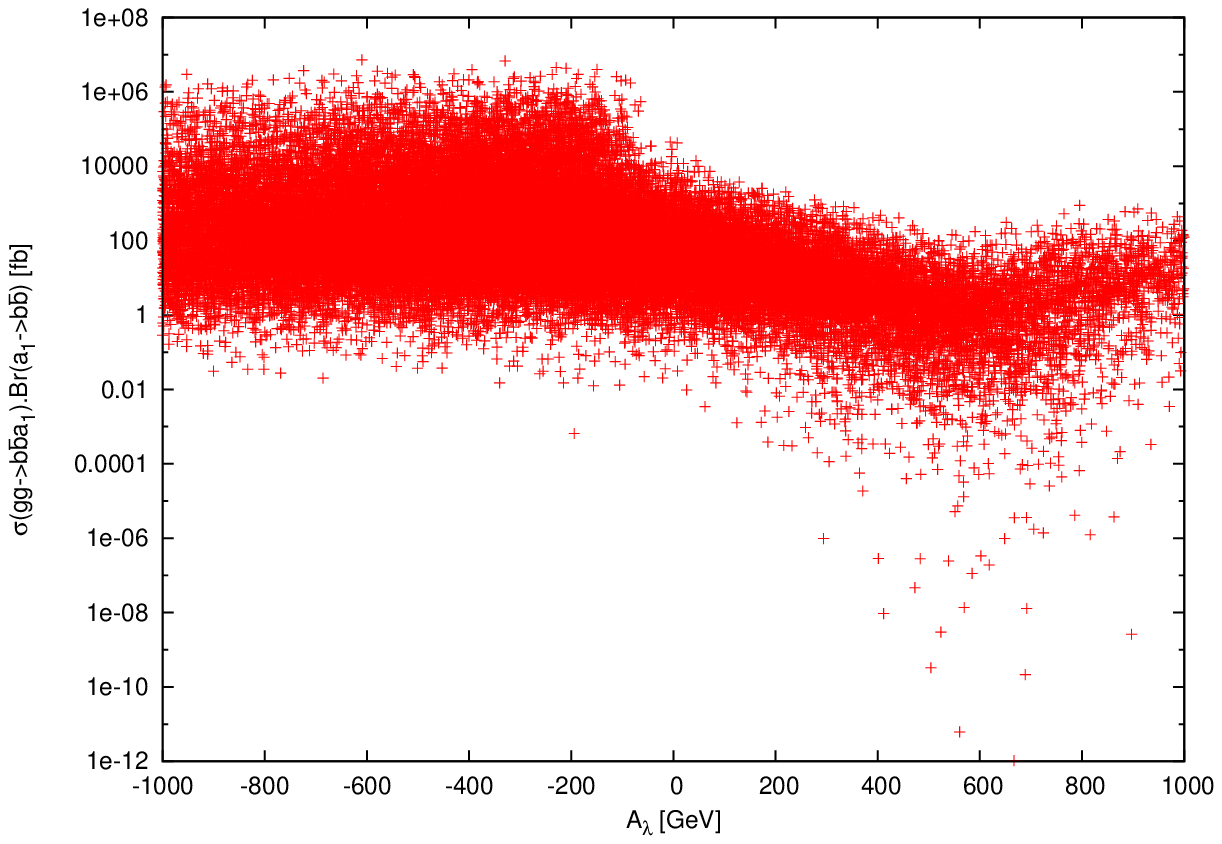}&\includegraphics[scale=0.60]{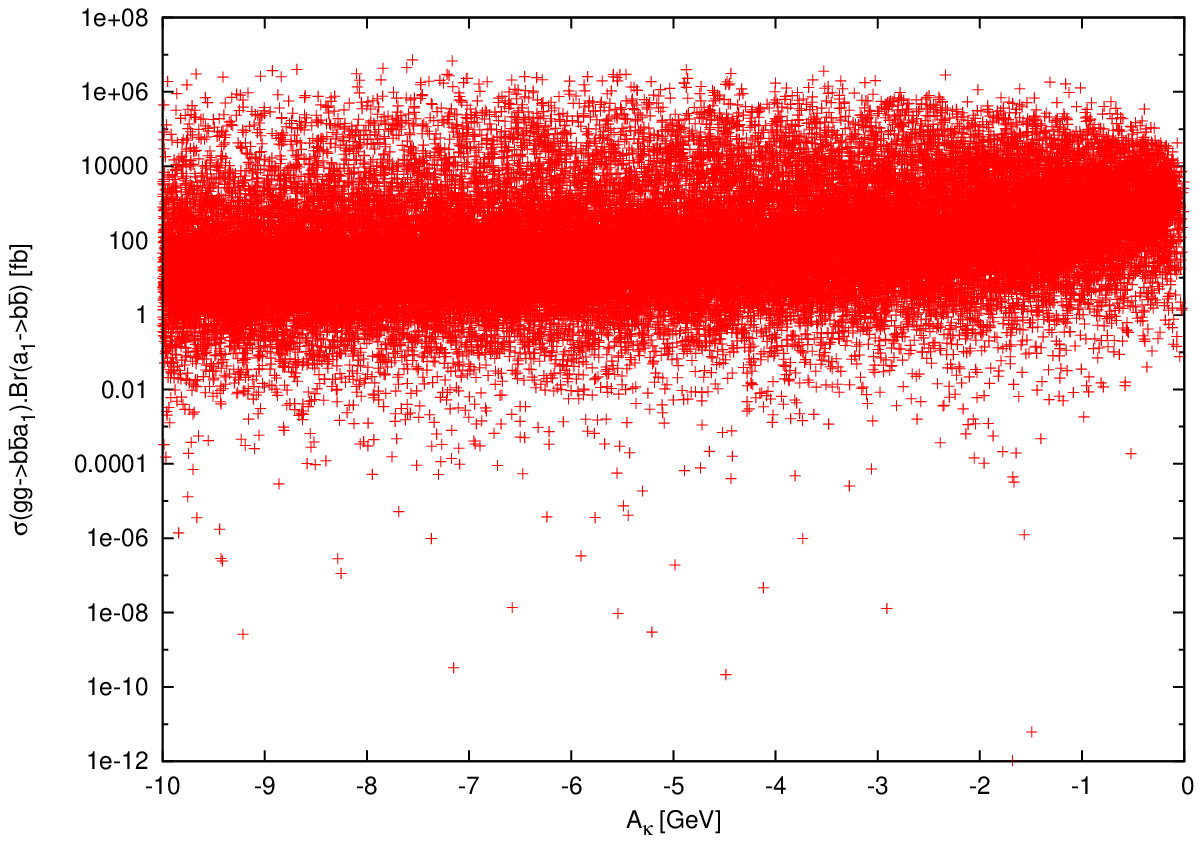}\\
     \includegraphics[scale=0.60]{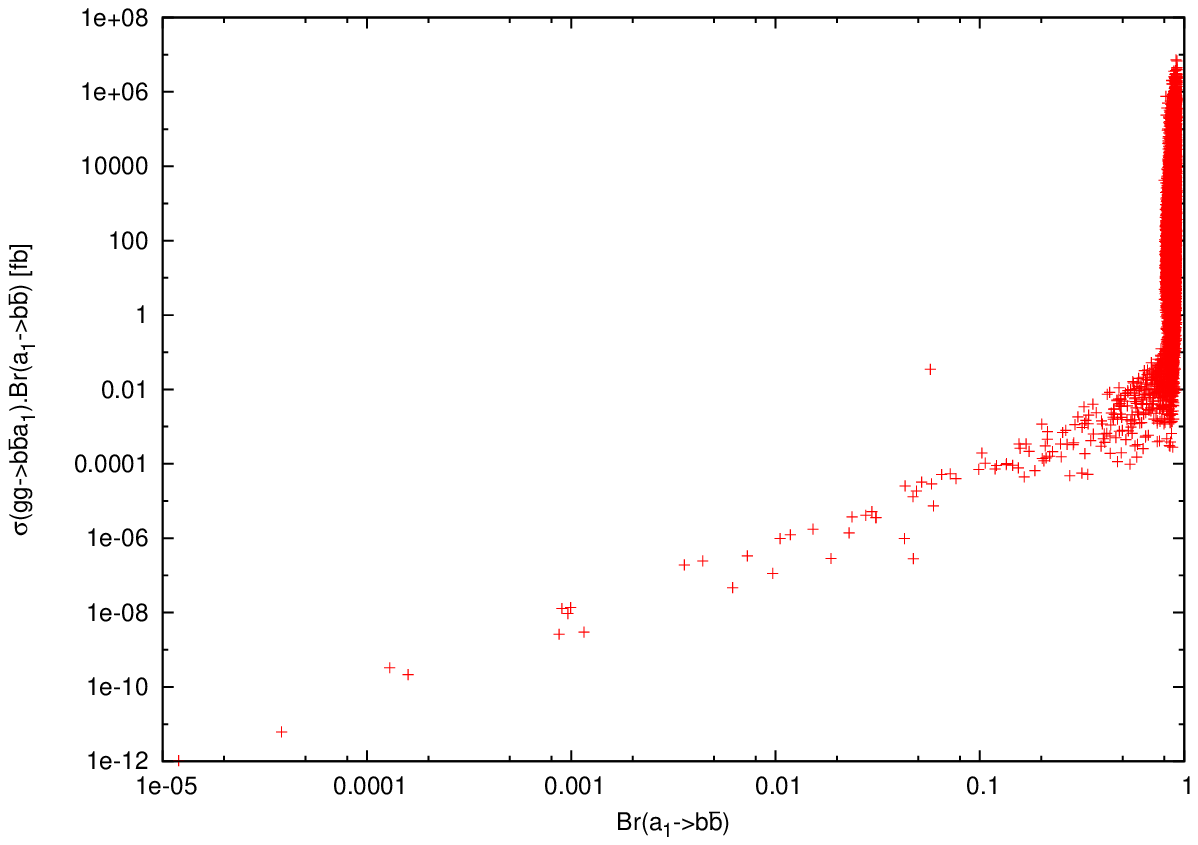}&\includegraphics[scale=0.60]{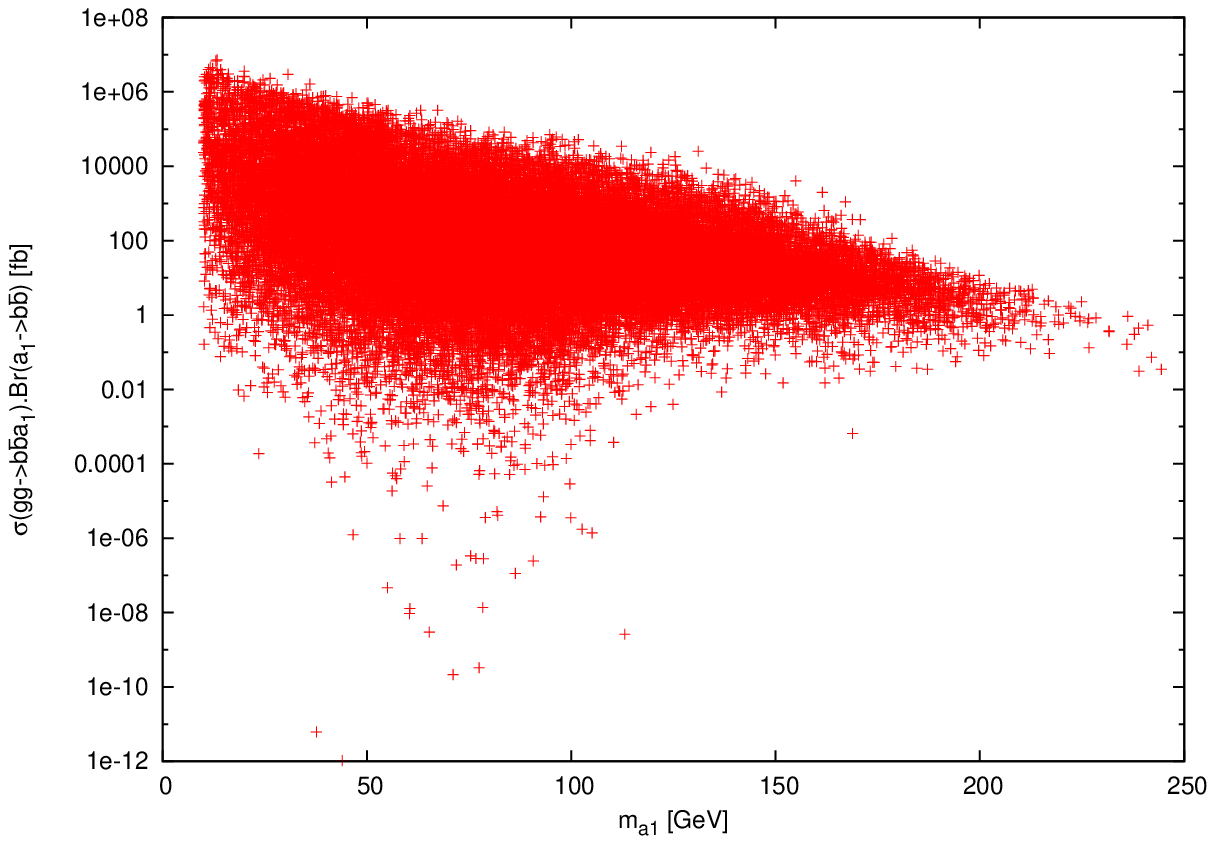}

 \end{tabular}
\label{fig:scan}
\caption{The rates for $\sigma(gg\to b\bar b {a_1})~{\rm BR}(a_1\to b\bar b)$ as a function of $\lambda$, 
$\kappa$, $\tan\beta$, $\mu_{\rm eff}$, $A_\lambda$, $A_\kappa$, Br$(a_1\to b\bar b)$ and $m_{a_1}$. }
\end{figure}

\begin{figure}
 \centering\begin{tabular}{cc}
 
   \includegraphics[scale=0.6]{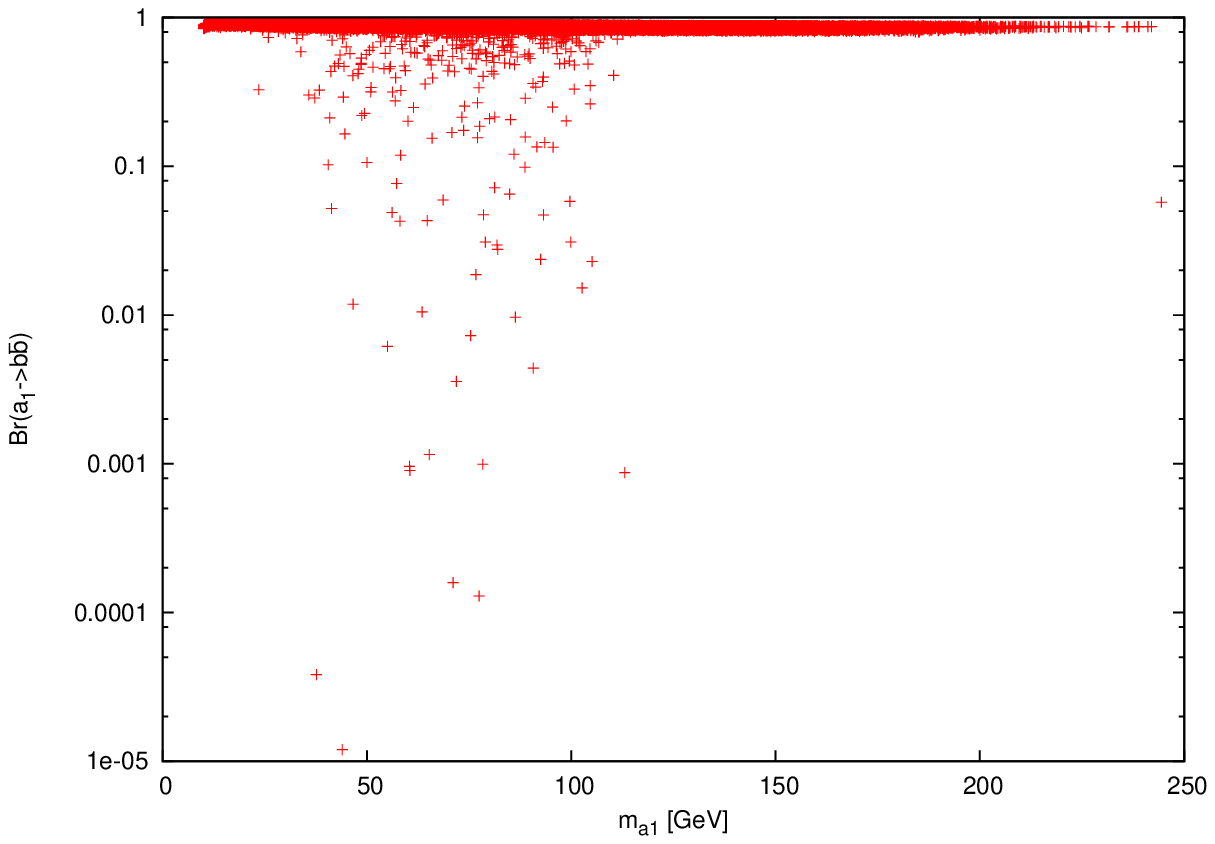}&\includegraphics[scale=0.6]{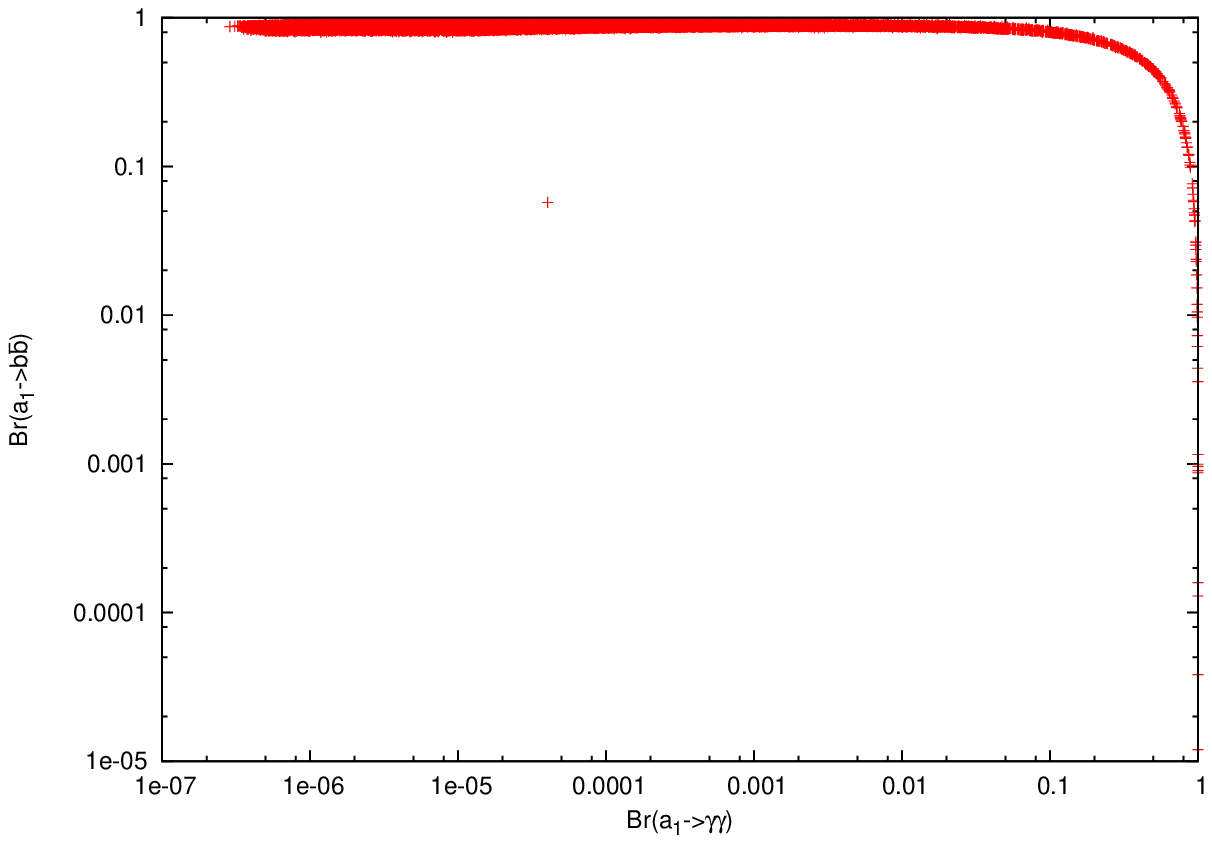}

 \end{tabular}
\label{fig:mass-branching ratio}
\caption{The BR$(a_1\to b\bar b)$ as a function of the CP-odd Higgs mass $m_{a_1}$ and of the BR$(a_1\to \gamma\gamma)$ .}
\end{figure}

\begin{figure}
 \centering\begin{tabular}{cc}
    \includegraphics[scale=0.7]{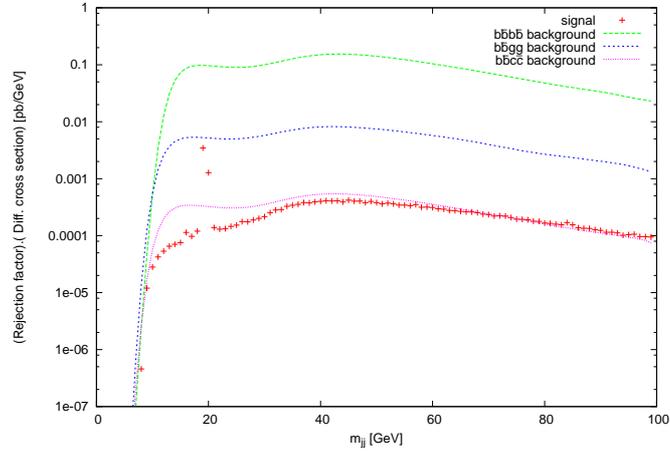}

 \end{tabular}
\label{fig:Mjj1}
\caption{The differential cross-section in di-jet invariant mass $m_{jj}$, after the cuts in (\ref{cuts:BB}), 
for signal (with $m_{a_1}$=19.98 GeV) and backgrounds, the former obtained for the benchmark point  
 with $\lambda$ = 0.075946278, $\kappa$ = 0.11543578, tan$\beta$ = 51.507125, $\mu$ = 377.4387,
 $A_\lambda$ = -579.63592 and $A_\kappa$ = -3.5282881.}

\end{figure}

\begin{figure}
 \centering\begin{tabular}{cc}
    \includegraphics[scale=0.7]{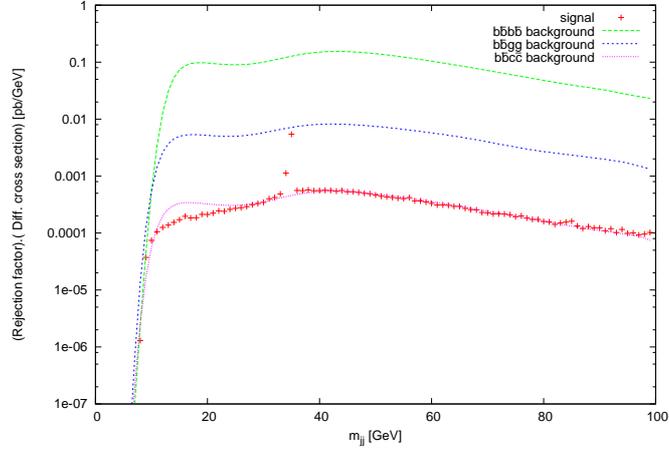}

 \end{tabular}
\label{fig:Mjj2}
\caption{The differential cross-section in di-jet invariant mass $m_{jj}$, after the cuts in (\ref{cuts:BB}),
for signal (with $m_{a_1}$=35.14 GeV) and backgrounds, the former obtained for the benchmark point  
 with $\lambda$ = 0.091741231, $\kappa$ = 0.51503049, tan$\beta$ = 38.09842, $\mu$ = 130.56601,
 $A_\lambda$ = -720.88387 and $A_\kappa$ = -5.3589352.}
\end{figure}

\begin{figure}
 \centering\begin{tabular}{cc}
   
     \includegraphics[scale=0.7]{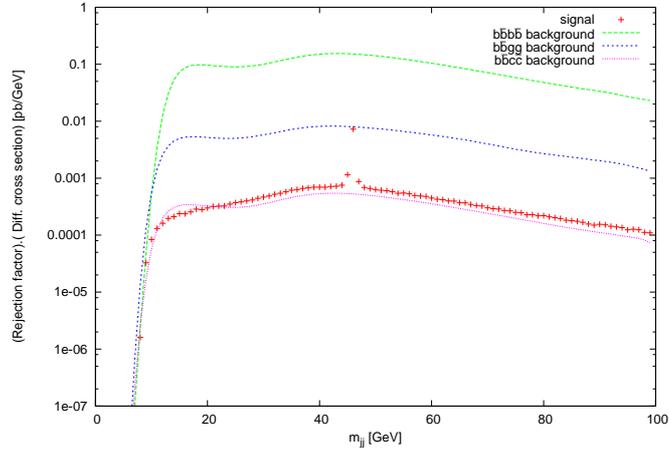}

 \end{tabular}
\label{fig:Mjj3}
\caption{The differential cross-section in di-jet invariant mass $m_{jj}$, after the cuts in (\ref{cuts:BB}),
 for signal (with $m_{a_1}$=46.35 GeV) and backgrounds, the former obtained for the benchmark point  
with $\lambda$ = 0.14088263, $\kappa$ = 0.25219468, tan$\beta$ = 50.558484, $\mu$ = 317.07532,
 $A_\lambda$ = -569.60665 and $A_\kappa$ = -8.6099538.}

\end{figure}

\begin{figure}
 \centering\begin{tabular}{cc}
    \includegraphics[scale=0.7]{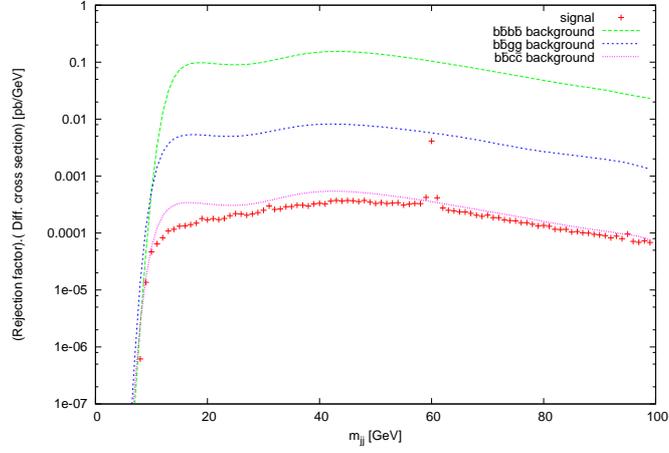}

 \end{tabular}
\label{fig:Mjj4}
\caption{The differential cross-section in di-jet invariant mass $m_{jj}$, after the cuts in (\ref{cuts:BB}),
 for signal (with $m_{a_1}$=60.51 GeV) and backgrounds, the former obtained for the benchmark point  
with $\lambda$ = 0.17410656, $\kappa$ = 0.47848034, tan$\beta$ = 52.385408, $\mu$ = 169.83139,
 $A_\lambda$ = -455.85097 and $A_\kappa$ = -9.0278415.}
\end{figure}

\begin{figure}
 \centering\begin{tabular}{cc}

     \includegraphics[scale=0.7]{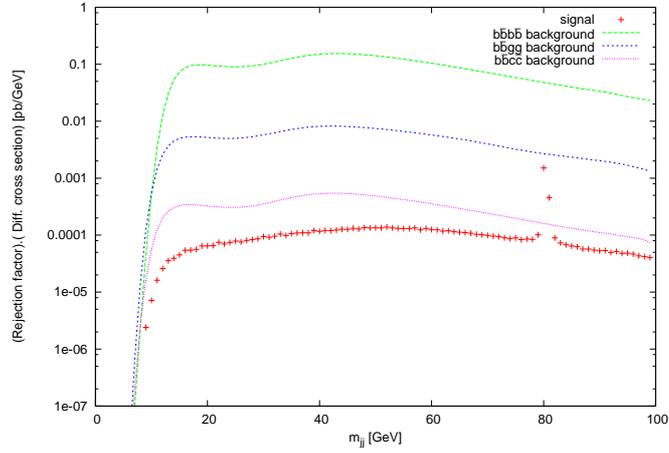}

 \end{tabular}
\label{fig:Mjj5}
\caption{The differential cross-section in di-jet invariant mass $m_{jj}$, after the cuts in (\ref{cuts:BB}),
 for signal (with $m_{a_1}$=80.91 GeV) and backgrounds, the former obtained for the benchmark point  
with $\lambda$ = 0.10713292, $\kappa$ = 0.13395171, tan$\beta$ = 44.721569, $\mu$ = 331.43456, $A_\lambda$ = -418.13018
 and $A_\kappa$ = -9.7077267.}
\end{figure}

\begin{figure}
 \centering\begin{tabular}{cc}
    \includegraphics[scale=0.6]{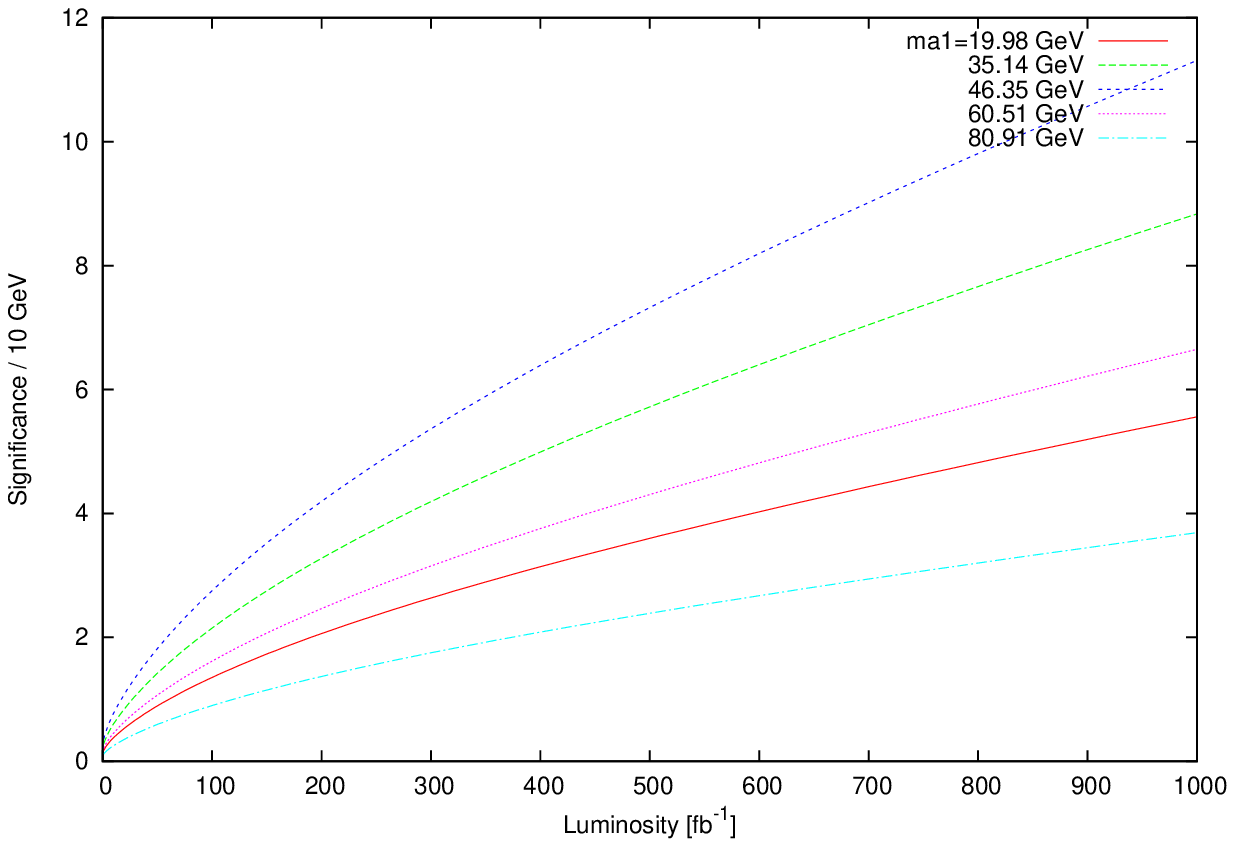}&\includegraphics[scale=0.6]{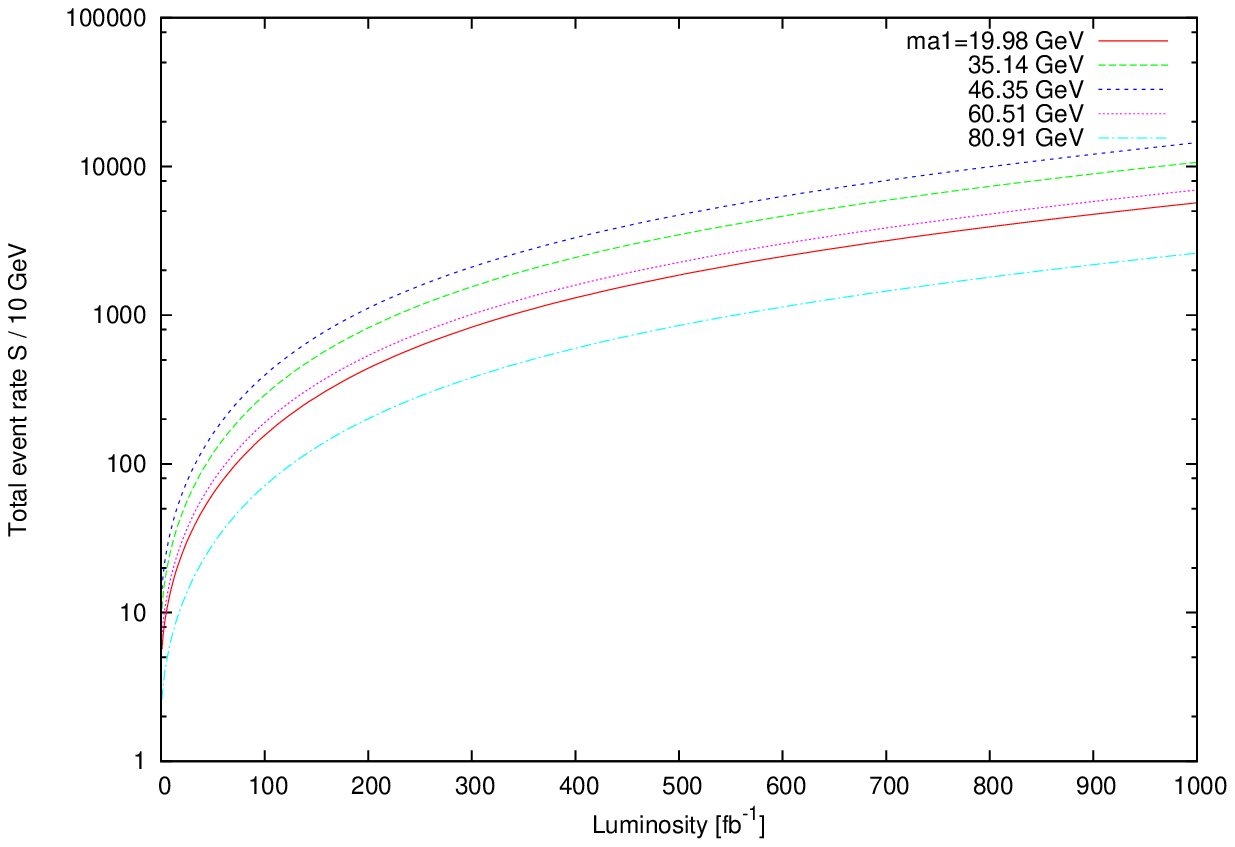}\\
     \includegraphics[scale=0.6]{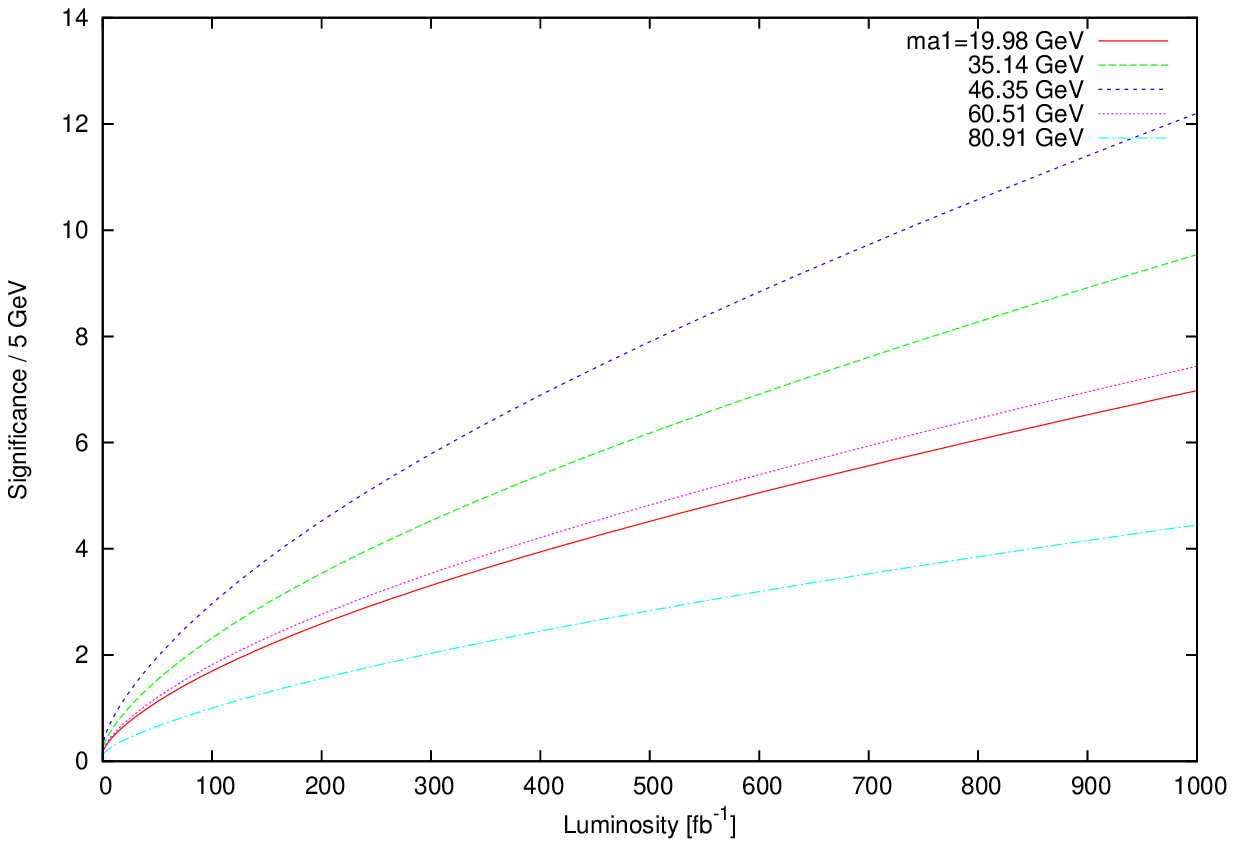}&\includegraphics[scale=0.6]{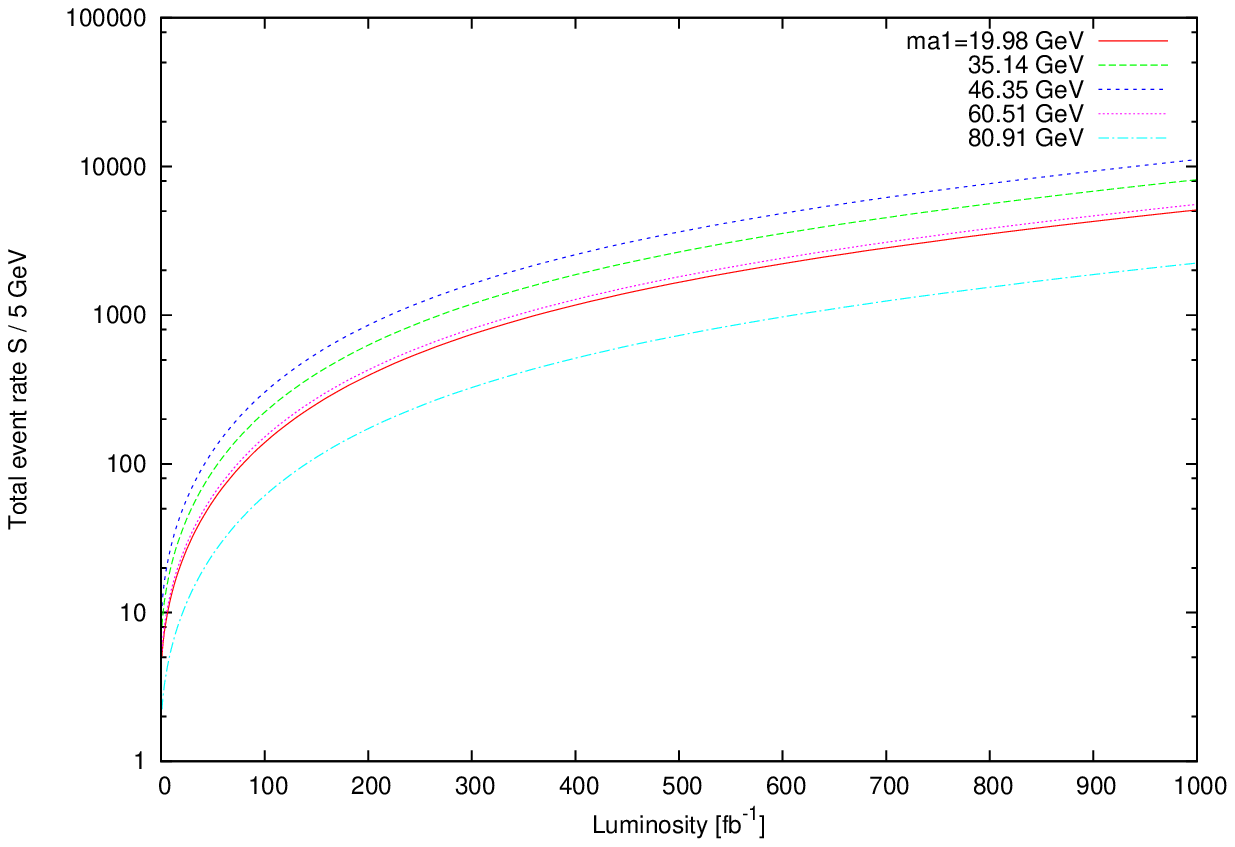}

 \end{tabular}
\label{fig:significance}
\caption{The significances $S/\sqrt B$ (left) and total event rates $S$ (right) of 
the $gg, q\bar q\to b\bar b a_1\to b\bar bb\bar b $ signal as functions 
of the integrated luminosity for 10 GeV (top) and 5 GeV (bottom)
di-jet mass resolutions.}
\end{figure}

\begin{figure}
 \centering\begin{tabular}{cc}
   
     \includegraphics[scale=0.6]{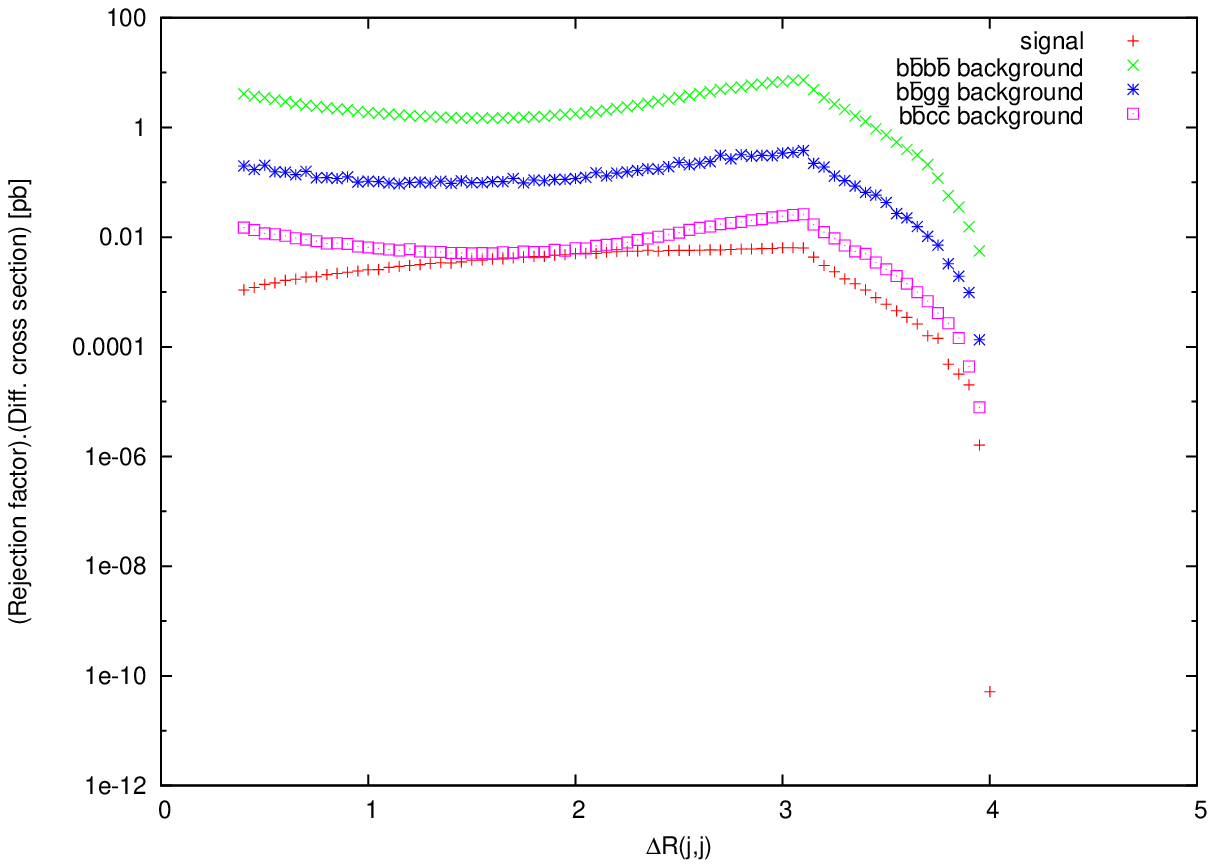}&\includegraphics[scale=0.6]{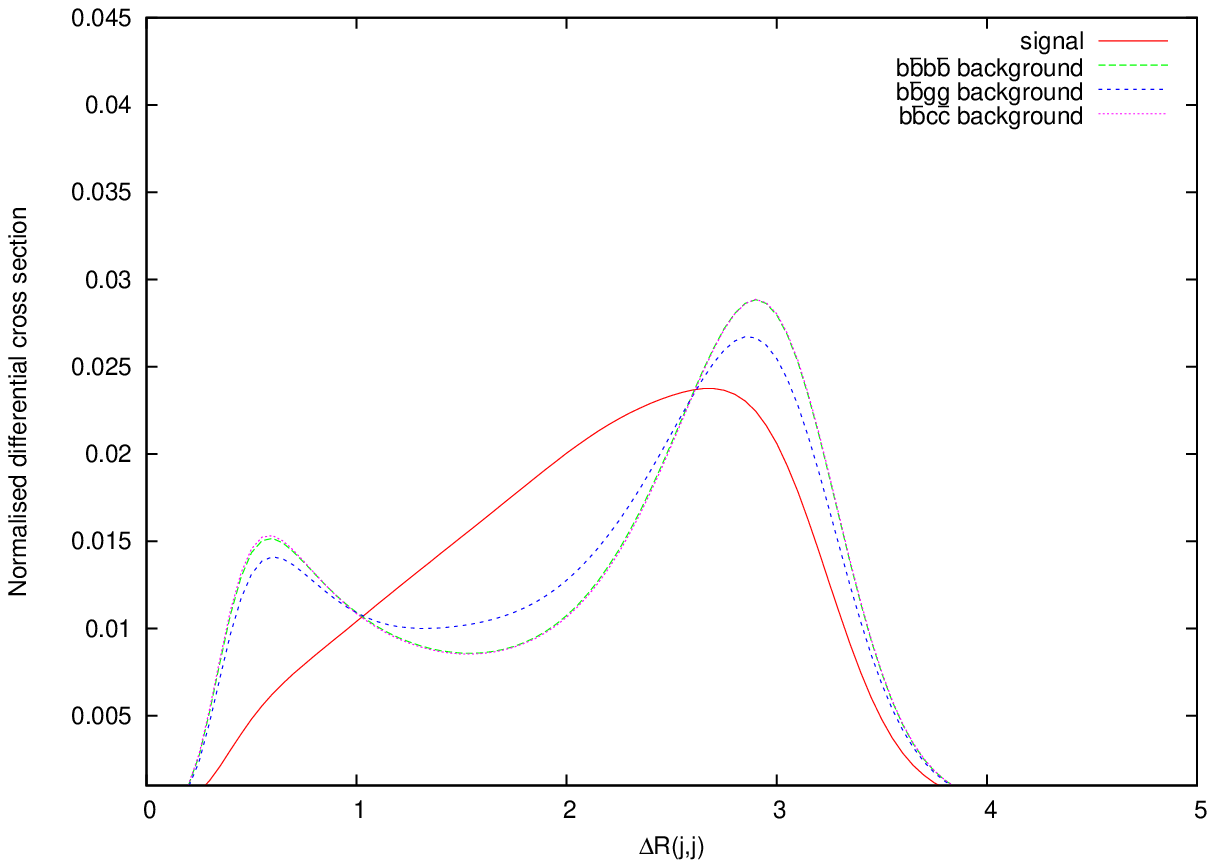}

 \end{tabular}
\label{fig:DeltaR}
 \caption{The differential cross-section in di-jet (pseudorapidity-azimuth) separation, after the cuts
 in (\ref{cuts:BB}), for signal (with $m_{a_1}$=46.35 GeV) and backgrounds, the former obtained for the benchmark point  
with $\lambda$ = 0.14088263, $\kappa$ = 0.25219468, tan$\beta$ = 50.558484, $\mu$ = 317.07532,
 $A_\lambda$ = -569.60665 and $A_\kappa$ = -8.6099538.
On the left the absolute normalisations are given while on the right those to 1.}

\end{figure}

\begin{figure}
 \centering\begin{tabular}{cc}
   
     \includegraphics[scale=0.6]{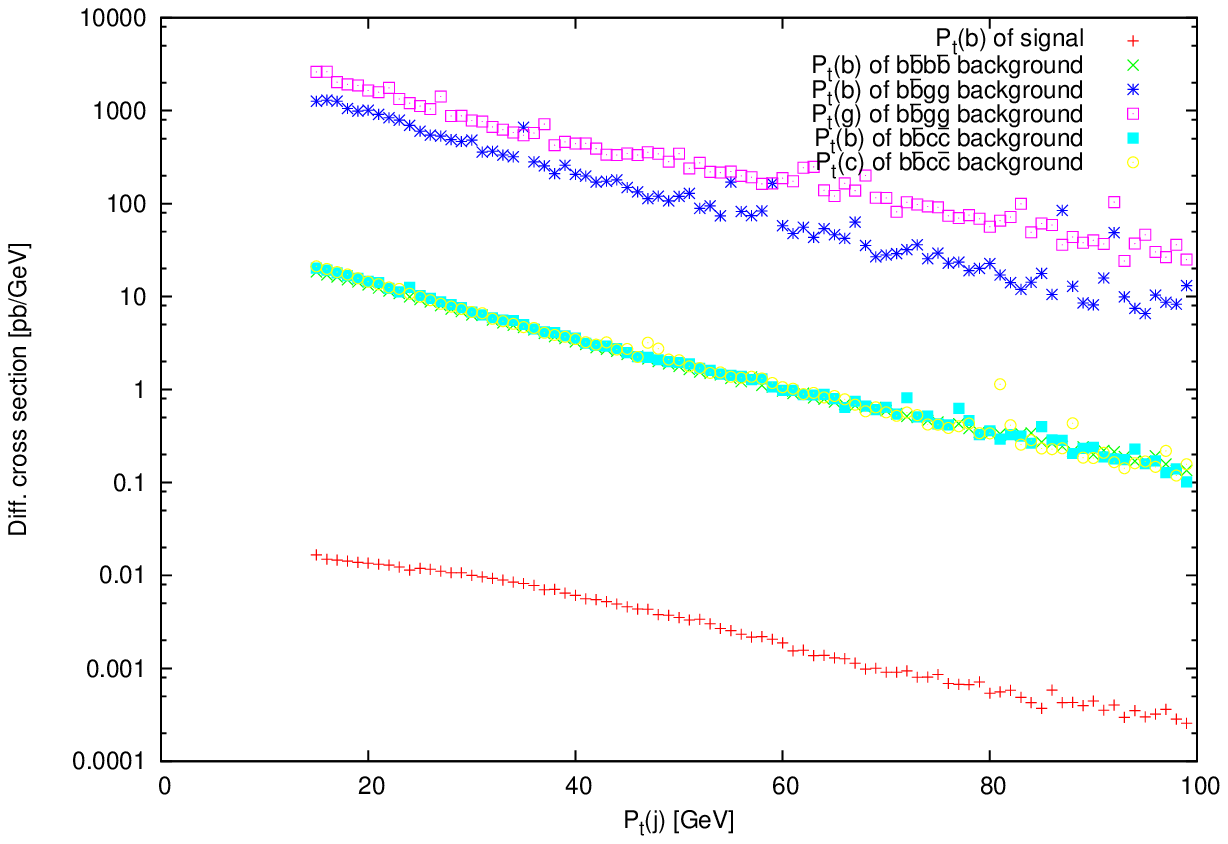}&\includegraphics[scale=0.6]{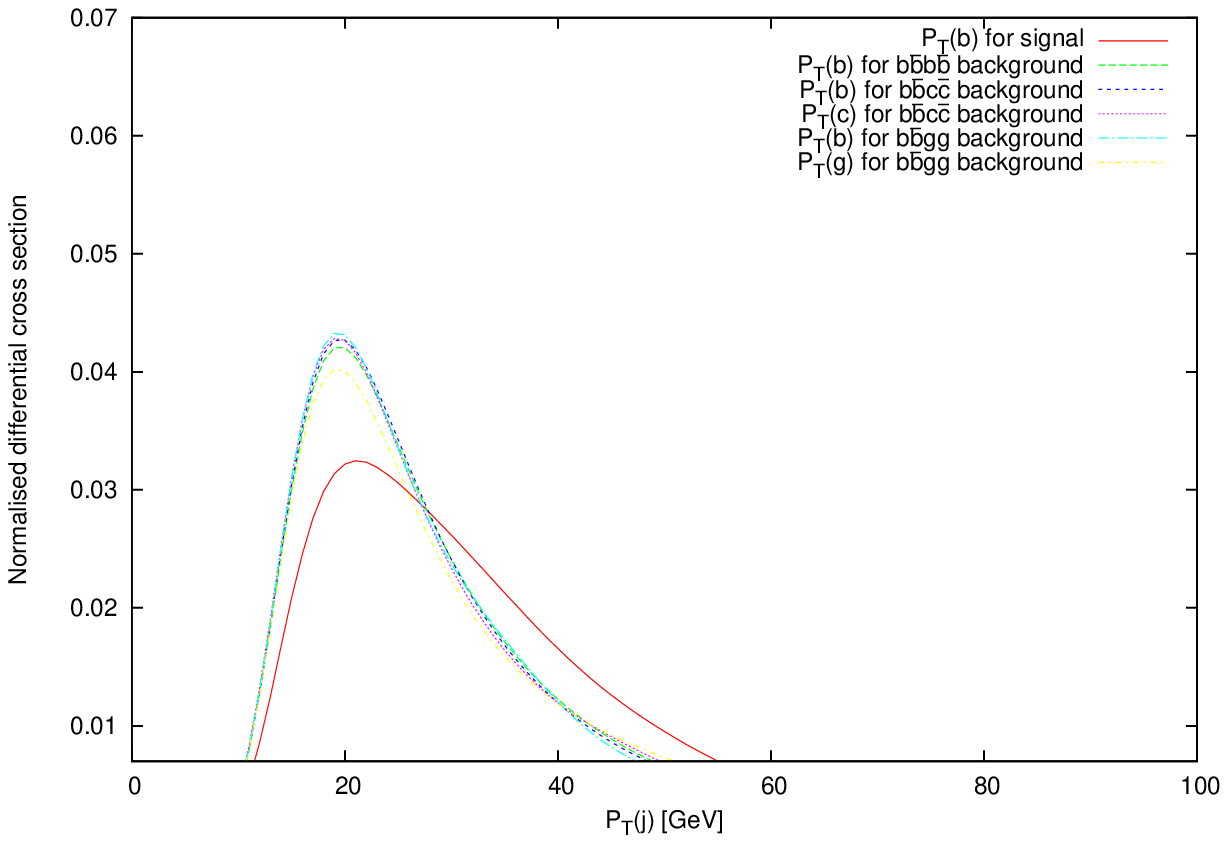}

 \end{tabular}
\label{fig:DeltaR}
\caption{The differential cross-section in jet transverse 
momentum, after the cuts in (\ref{cuts:BB}), for signal (with $m_{a_1}$=46.35 GeV) and backgrounds,
the former obtained for the benchmark point  
with $\lambda$ = 0.14088263, $\kappa$ = 0.25219468, tan$\beta$ = 50.558484, $\mu$ = 317.07532,
 $A_\lambda$ = -569.60665 and $A_\kappa$ = -8.6099538.
On the left the absolute normalisations are given while on the right those to 1.}

\end{figure}

\end{document}